\documentclass[conference]{IEEEtran}
\IEEEoverridecommandlockouts
\usepackage{cite}
\usepackage{amsmath,amssymb,amsfonts}
\usepackage{algorithmic}
\usepackage{graphicx}
\usepackage{textcomp}
\usepackage{xcolor}
\usepackage{listings}
\usepackage{subcaption}
 
\definecolor{codegreen}{rgb}{0,0.6,0}
\definecolor{codegray}{rgb}{0.5,0.5,0.5}
\definecolor{codepurple}{rgb}{0.58,0,0.82}
\definecolor{backcolour}{rgb}{0.95,0.95,0.92}
 
\def\BibTeX{{\rm B\kern-.05em{\sc i\kern-.025em b}\kern-.08em
    T\kern-.1667em\lower.7ex\hbox{E}\kern-.125emX}}

\begin{document}

\title{Vortex: OpenCL Compatible RISC-V GPGPU
}

\author{\IEEEauthorblockN{Fares Elsabbagh}
\IEEEauthorblockA{\textit{Georgia Tech}\\
fsabbagh@gatech.edu}
\and
\IEEEauthorblockN{Blaise Tine}
\IEEEauthorblockA{\textit{Georgia Tech}\\
btine3@gatech.edu}
\and
\IEEEauthorblockN{Priyadarshini Roshan}
\IEEEauthorblockA{\textit{Georgia Tech}\\
priya77darshini@gatech.edu}
\and
\IEEEauthorblockN{Ethan Lyons}
\IEEEauthorblockA{\textit{Georgia Tech}\\
elyons8@gatech.edu}
\and
\IEEEauthorblockN{Euna Kim}
\IEEEauthorblockA{\textit{Georgia Tech}\\
ekim79@gatech.edu}
\and
\IEEEauthorblockN{Da Eun Shim}
\IEEEauthorblockA{\textit{Georgia Tech}\\
daeun@gatech.edu}
\and
\IEEEauthorblockN{Lingjun Zhu}
\IEEEauthorblockA{\textit{Georgia Tech}\\
lingjun@gatech.edu}
\and
\IEEEauthorblockN{Sung Kyu Lim}
\IEEEauthorblockA{\textit{Georgia Tech}\\
limsk@ece.gatech.edu}
\and
\IEEEauthorblockN{Hyesoon Kim}
\IEEEauthorblockA{\textit{Georgia Tech}\\
hyesoon@cc.gatech.edu}\\
}

\maketitle

\begin{abstract}
\label{sec:abstract}

The current challenges in technology scaling are pushing the semiconductor industry towards hardware specialization, creating a proliferation of heterogeneous systems-on-chip, delivering orders of magnitude performance and power benefits compared to traditional general-purpose architectures. This transition is getting a significant boost with the advent of RISC-V with its unique modular and extensible ISA, allowing a wide range of low-cost processor designs for various target applications. In addition, OpenCL is currently the most widely adopted programming framework for heterogeneous platforms available on mainstream CPUs, GPUs, as well as FPGAs and custom DSP.

In this work, we present Vortex, a RISC-V General-Purpose GPU that supports OpenCL. Vortex implements a SIMT architecture with a minimal ISA extension to RISC-V that enables the execution of OpenCL programs. We also extended OpenCL runtime framework to use the new ISA.  We evaluate this design using  15nm technology. We also show the performance and energy numbers of running them with a subset of benchmarks from the Rodinia Benchmark suite.

\end{abstract}

\begin{IEEEkeywords}
GPGPU, OpenCL, Vector processors 
\end{IEEEkeywords}

\section{Introduction}
\label{sec:introduction}

The emergence of data parallel architectures and general purpose graphics processing units (GPGPUs) have enabled new opportunities to address the power limitations and scalability of multi-core processors, allowing new ways to exploit the abundant data parallelism present in emerging big-data parallel applications such as machine learning and graph analytics. GPGPUs in particular, with their Single Instruction Multiple-Thread (SIMT) execution model, heavily leverage data-parallel multi-threading to maximize throughput at relatively low energy cost, leading the current race for energy efficiency (Green500 \cite{Green500}) and applications support with their accelerator-centric parallel programming model (CUDA \cite{CUDA} and OpenCL \cite{OpenCL}).

The advent of RISC-V~\cite{waterman2011risc, waterman2014risc, asanovic2014instruction}, open-source and free instruction set architecture (ISA), provides a new level of freedom in designing hardware architectures at lower cost, leveraging its rich eco-system of open-source software and tools. With RISC-V, computer architects have designed several innovative processors and cores such as BOOM~v1 and BOOM~v2~\cite{celio2017boomv2} out-of-order cores, as well as  system-on-chip (SoC) platforms for a wide range of applications. For instance, Gautschi et al. \cite{gautschi2017near} have extended RISC-V to digital signal processing (DSP) for scalable Internet-of-things (IoT) devices. Moreover, vector processors \cite{zimmer2015risc} \cite{HWACHA} \cite{ARA} and processors integrated with vector accelerators \cite{lee201445nm} \cite{MANIC} have been designed and fabricated based on RISC-V. In spite of the advantages of the preceding works, not enough attention has been devoted to building an open-source general-purpose GPU (GPGPU) system based on RISC-V.

Although a couple of recent work have been proposed for massively parallel computations on FPGA using RISC-V, (GRVI Phalanx) \cite{gray2016grvi}, (Simty) \cite{collange2017simty}, none of them have implemented the full-stack by extending the RISC-V ISA, synthesizing the microarchitecture, and implementing the software stack to execute OpenCL programs. We believe that such an implementation is in fact necessary to achieve the level of usability and customizability in massively parallel platforms.

In this paper, we propose an ISA RISC-V extension for GPGPU programs and microarchitecture. We also extend a software stack to support OpenCL.


This paper makes the following key contributions:
\begin{itemize}
\item We propose a highly configurable SIMT-based General Purpose GPU architecture targeting the RISC-V ISA and synthesized the design using a Synopsys library with our RTL design. 
\item We show that the minimal set of five instructions on top of RV32IM (RISC-V 32 bit integer and multiply extensions) enables SIMT execution.  
\item We describe the necessary changes in the software stack that enable the execution of OpenCL programs on Vortex. We demonstrate the portability by running a subset of Rodinia benchmarks~\cite{rodinia_bench}. 
\end{itemize}


\begin{figure}[t]
\centering
\includegraphics[width=0.5\textwidth]{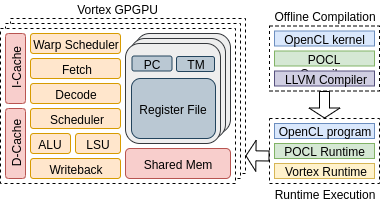}
\caption{Vortex System Overview}
\label{fig:sw}
\end{figure}

\section{Background}
\label{sec:background}

\subsection{Open-Source OpenCL Implementations} \label{os_opencl_imp}


POCL \cite{POCL} implements a flexible compilation backend based on LLVM, allowing it to support a wider range of device targets including general purpose processors (e.g. x86, ARM, Mips), General Purpose GPU (e.g. Nvidia), and TCE-based processors \cite{TCE} and custom accelerators. The custom accelerator support provides an efficient solution for enabling OpenCL applications to use hardware devices with specialized fixed-function hardware (e.g. SPMV, GEMM). POCL is comprised of two main components: A back-end compilation engine, and a front-end OpenCL runtime API.

The POCL runtime implements the device-independent common interface where the target implementation of each device plugs into POCL to specialize their operations. At runtime, POCL invokes the back-end compiler with the provided OpenCL kernel source. POCL supports target-specific execution models including SIMT, MIMD, SIMD, and VLIW. On platforms supporting MIMD and SIMD execution models such as CPUs, the POCL compiler attempts to pack as many OpenCL work-items to the same vector instruction, then the POCL runtime will distribute the remaining work-items among the active hardware threads on the device with provided synchronization. On platforms supporting SIMT execution model such as GPUs, the POCL compiler delegates the distribution of the work-items to the hardware to spread the execution among its hardware threads, relying on the device to also handle the necessary synchronization. On platforms supporting VLIW execution models such as TCE-based accelerators, the POCL compiler attempts to "unroll" the parallel regions in the kernel code such that the operations of several independent work-items can be statically scheduled to the multiple function units of the target device.
\section{The OpenCL Software Stack}
\label{sec:software}

\begin{figure}
\centering
\includegraphics[width=0.5\textwidth]{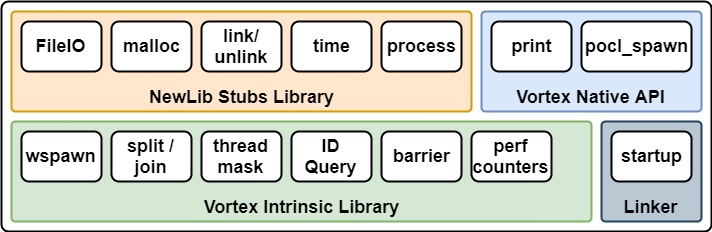}
\caption{Vortex Runtime Library}
\label{fig:runtime}
\end{figure}

\begin{figure}
  \centering
\begin{lstlisting}
vx_intrinsic.s 
vx_split:
	.word 0x0005206b    # split a0
	ret
vx_join:
	.word 0x0000306b    #join
	ret

kernel.cl 
#define __if(cond) split(cond); \
          if(cond)
          
#define __endif join();
void opencl_kernel(){
    int id = vx_getTid();
    __if(id<4) {
        // Path A
    } else {
        // Path B
    } __endif
}
         

\end{lstlisting}
  \caption{This figure shows the control divergent \_\_if \_\_endif macro definitions and how they could be used to enable control divergence in OpenCL kernels. Currently, this process is done manually for each kernel.}
  \label{fig:ifelseimpl}
\end{figure}

\subsection{Vortex Native Runtime} \label{vx_runtime}

The Vortex software stack implements a native runtime library for developing applications that will run on Vortex and take advantage of the new RISC-V ISA extension. Figure \ref{fig:runtime} illustrates the Vortex runtime layer, which is comprised of three main components: 1) Low-level intrinsic library exposing the new ISA interface, 2) A support library that implements NewLib stub functions \cite{Newlib}, 3) a native runtime API for launching POCL kernels.



\subsubsection{Instrinsic Library}
To enable Vortex runtime kernel to utilize the new instructions without modifying the existing compilers, we implemented an intrinsic  layer that implements the new ISA. Figure~\ref{fig:runtime} shows the functions and ISA supported by the intrinsic library. We leverage RISC-V's ABI which guarantees function arguments being passed through the argument registers and return values begin passed through \textit{a0} register. Thus, these intrinsic functions have only two assembly instructions: 1) The encoded 32-bit hex representation of the instruction that uses the argument registers as source registers, and 2) a return instruction that returns back to the C++ program. An example of these intrinsic functions is illustrated in Figure~\ref{fig:ifelseimpl}. In addition, to handle control divergence, which is frequent in OpenCL kernels, we implement \_\_if and \_\_endif macros shown in Figure~\ref{fig:ifelseimpl} to handle the insertion of these intrinsic functions with minimal changes to the code. These changes are currently done manually for the OpenCL kernels. This approach achieves the required functionality without restricting the platform or requiring any modifications to the RISC-V compilers.

\subsubsection{Newlib Stubs Library}
Vortex software stack uses the NewLib \cite{Newlib} library to enable programs to use the C/C++ standard library without the need to support an operating system. NewLib defines a minimal set of stub functions that client applications need to implement to handle necessary system calls such as file I/O, allocation, time, process, etc.. 


\subsubsection{Vortex Native API}
\begin{figure}
\centering
\includegraphics[width=0.5\textwidth]{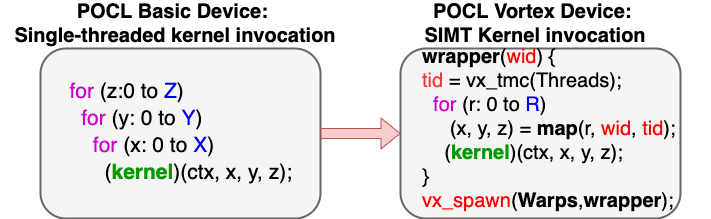}
\caption{SIMT Kernel invocation modification for Vortex in POCL} 
\label{fig:pocl_spawn}
\end{figure}

The Vortex native API implements some general purpose utility routines for applications to use. One of such routines is \textit{pocl\_spawn()} which allows programs to schedule POCL kernel execution on Vortex. \textit{pocl\_spawn()} is responsible for mapping work groups requested by POCL to the hardware: 1) It uses the intrinsic layer to find out the available hardware resources, 2) Uses the requested work group dimension and numbers to divide the work equally among the hardware resources, 3) For each OpenCL dimension, it assigns a range of IDs to each available warp in a global structure, 4) It uses the intrinsic layer to spawn the warps and activate threads, and finally 5) Each warp will loop through the assigned IDs, executing the kernel every time with a new OpenCL \textit{global\_id}. Figure~\ref{fig:pocl_spawn} shows an example. In the original OpenCL code, the \textit{kernel} is called once with global/local sizes as arguments. POCL wraps the kernel with three loops and sequentially calls with the logic that converts x,y,z to global ids. For a Vortex version, warps and threads are spawned, then each thread is assigned a different work-group to execute the kernel. POCL provides the feature to map the correct \textit{wid}, which was a part of the baseline POCL implementation to support various hardware such as vector architecture.


\subsection{POCL Runtime}

We modified the POCL runtime, adding a new device target to its common device interface to support Vortex. The new device target is essentially a variant of the POCL basic CPU target with the support for pthreads and other OS dependencies removed to target the NewLib interface. We also modified the single-threaded logic for executing work-items to use Vortex's \textit{pocl\_spawn} runtime API.


\subsection{Barrier Support}

Synchronizations within work-group in OpenCL is supported by barriers. POCL's back-end compiler splits the control-flow graph (CFG) of the kernel around the barrier and splits the kernel into two sections that will be executed by all local work-groups sequentially.

\section{Vortex Parallel Hardware Architecture}
\label{sec:hardware}

\begin{figure*}[t]
\centering
\includegraphics[width=1.5\columnwidth]{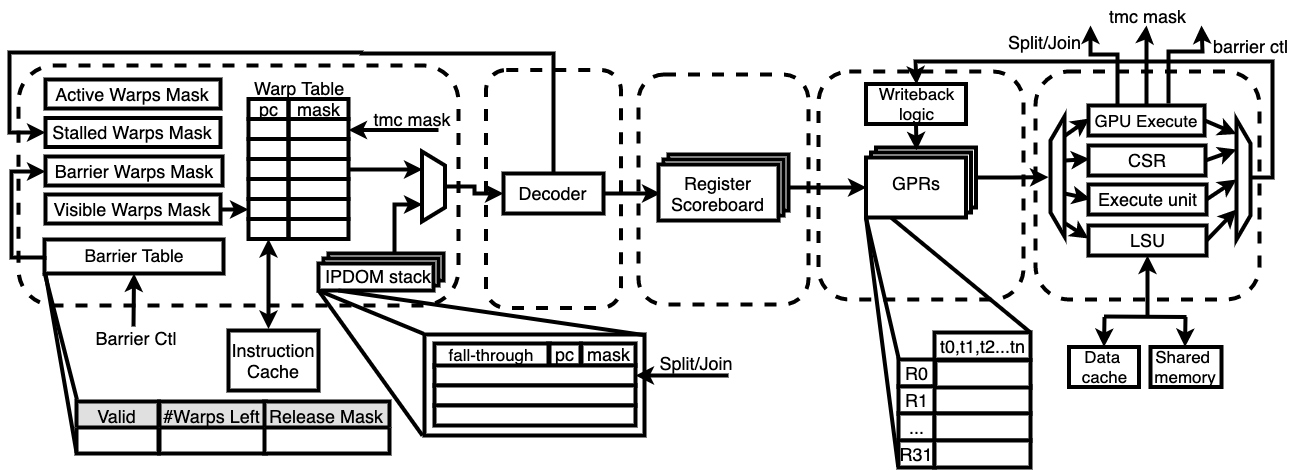}
\caption{Vortex Microarchitecture.}
\label{fig:VortexMicro}
\end{figure*}

\begin{figure}
\centering
\includegraphics[scale=0.3]{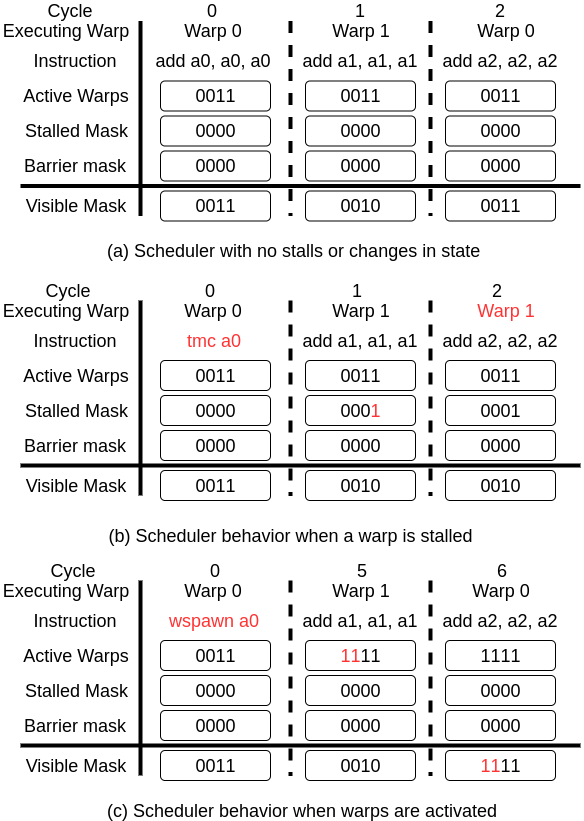}
\caption{This figure shows the Warp Scheduler under different scenarios. In the actual microarchitecture implementation, the instruction is only known the next cycle, however it's displayed in the same cycle in this figure for simplicity.}
\label{fig:warpSched}
\end{figure}




\subsection{SIMT Hardware Primitives} \label{hw_primitives}

SIMT, Single Instruction-Multiple Threads, execution model takes advantage of the fact that in most parallel applications, the same code is repeatedly executed but with different data. Thus, it provides the concept of Warps~\cite{CUDA}, which is a group of threads that share the same PC and follows the same execution path with minimal divergence. Each thread in a warp has a private set of general purpose registers, and the width of the ALU is matched with the number of threads. However, the fetching, decoding, and issuing of instructions is shared within the same warp which reduces execution cycles.

However, in some cases, the threads in the same warp will not agree on the direction of branches. In such cases, the hardware must provide a thread mask to predicate instructions for each thread, and an IPDOM stack, to ensure all threads execute correctly, which are explained in Section~\ref{hw_threads}.


\subsection{Warp Scheduler}

The warp scheduler is in the fetch stage which decides what to fetch from I-cache as shown in Figure~\ref{fig:VortexMicro}. It has two components: 1) A set of warp masks to choose the warp to schedule next, and 2) a warp table that includes private information for each warp.

There are 4 thread masks that the scheduler uses: 1) an active warps mask, one bit indicating whether a warp is active or not, 2) a stalled warp mask, which indicates which warps should not be scheduled temporarily ( e.g., waiting for a memory request), 3) a barrier warps stalled mask, which indicates warps that have been stalled because of a barrier instruction, and 4) a visible warps mask to support hierarchical scheduling policy~\cite{nargpu11}.

Each cycle, the scheduler selects one warp from the visible warp mask and invalidates that warp. When visible warp mask is zero, the active mask is refilled by checking which warps are currently active and not stalled. 


An example of the warp scheduler is shown in Figure~\ref{fig:warpSched}(a). This figure shows the normal execution; The first cycle warp zero executes an instruction, then in the second cycle warp zero is invalidated in the visible warp mask, and warp one is scheduled. In third cycle, because there are no more warps to be scheduled, the scheduler uses the active warps to refill the visible mask, and schedules warp zero for execution.

Figure~\ref{fig:warpSched}(b) shows how the warp scheduler handles a stalled warp. In the first cycle the scheduler schedules warp zero. In the second cycle, the scheduler schedules warp one. At the same cycle,  because the decode stage identified that warp zero's instruction requires a change of state, it stalls warp zero. In the third cycle, because warp zero is stalled, the scheduler only sets warp one to visible warp mask and  schedules warp one again. When warp zero updates its thread mask, the bit in stalled mask will be set to 0 to allow scheduling.

Figure~\ref{fig:warpSched}(c) shows an example of spawning warps. When warp zero executes a \textit{wspawn} instruction (Table~\ref{table:isa}) , which activates warps and manipulates the active warps mask by setting warps two and three to be active. When it's time to refill the visible mask, because it  no longer has any warps to schedule, it includes warps two and three. Warps will stay in the Active Mask until they set their thread mask's value to zero, or warp zero utilizes wspawn to deactivate these warps.

\subsection{Threads Masks and IPDOM Stack} \label{hw_threads}

To support the thread concepts provided in Section~\ref{hw_primitives}, a thread mask register and an IPDOM stack have been added to the hardware similar to other SIMT architectures~\cite{Fung:2007:DWF:1331699.1331735}. The thread mask register acts like a predicate for each thread, controlling which threads are active. If the bit in the thread mask for a specific thread is zero, no modifications would be made to that thread's register file and no changes to the cache would be made based on that thread.

The IPDOM stack, illustrated in Figure~\ref{fig:VortexMicro} is used to handle control divergence and is controlled by the split and join instructions. These instructions utilize the IPDOM stack to enable divergence as shown in Figure~\ref{fig:ifelseimpl}.

When a split instruction is executed by a warp, the predicate value for each thread is evaluated. If there is only one thread active, or all threads agree on a direction, the split acts like a nop instruction and does not change the state of the warp. When there is more than one active thread that contradicts on the value of the predicate, three microarchitecture events occur: 1) The current thread mask is pushed into the IPDOM stack as a fall-through entry, 2) The active threads that evaluate the predicate as false are pushed into the stack with PC+4 (i.e., size of instruction) of the split instruction, and 3) The current thread mask is updated to reflect the active threads that evaluate the predicate to be true. 

When a join instruction is executed, an entry is popped out of the stack which causes one of two scenarios: 1) If the entry is not a fall-through entry, the PC is set to the entry's PC and the thread mask is updated to the value of the entry's mask, which enables the threads evaluating the predicate as false to follow their own execution path, and 2) If the entry is a fall-through entry, the PC continues executing to PC+4 and the thread mask is updated to the entry's mask, which is the case when both paths of the control divergence have been executed.

\subsection{Warp Barriers} \label{hw_barriers}

Warp barriers are important in SIMT execution, as it provides synchronization between warps. Barriers are provided in the hardware to support global synchronization between work-groups. Each barrier has a private entry in barrier table, shown in Figure~\ref{fig:VortexMicro}, with the following information: 1) Whether that barrier is currently valid, 2) the number of warps left that need to execute the barrier instruction with that entry's ID for the barrier to be released, and 3) a mask of the warps that are currently stalled by that barrier. However, Figure~\ref{fig:VortexMicro} only shows the per core barriers. There is also another table on multi-core configurations that allows for global barriers between all the cores. The MSB of the barrier ID indicates whether the instruction uses local barriers or global barriers.

When a barrier instruction is executed, the microarchitecture checks  the number of warps executed with the same barrier ID. If the number of warps is not equal to one, the warp is stalled until that number is reached and the release mask is manipulated to include that warp. Once the same number of warps have been executed, the release mask is used to release all the warps stalled by the corresponding barrier ID. The same method works for both local and global barriers; however, global barrier tables have a release mask per each core.

\begin{table}[h!]
\centering
\caption{Proposed SIMT ISA extension.}
\begin{tabular}{||c | c||} 
 \hline
 Instructions & Description \\ [0.5ex] 
 \hline\hline
 wspawn \%numW, \%PC & Spawn W new warps at PC  \\ 
 tmc \%numT & Change the thread mask to activate threads\\
 split \%pred & Control flow divergence \\
 join & Control flow reconvergence \\ [1ex] 
 bar \%barID, \%numW & Hardware Warps Barrier \\
 \hline
\end{tabular}
\label{table:isa}
\end{table}

\begin{figure}[h]
\centering
\begin{subfigure}{.48\columnwidth}
\centering
  \includegraphics[width=.65\columnwidth]{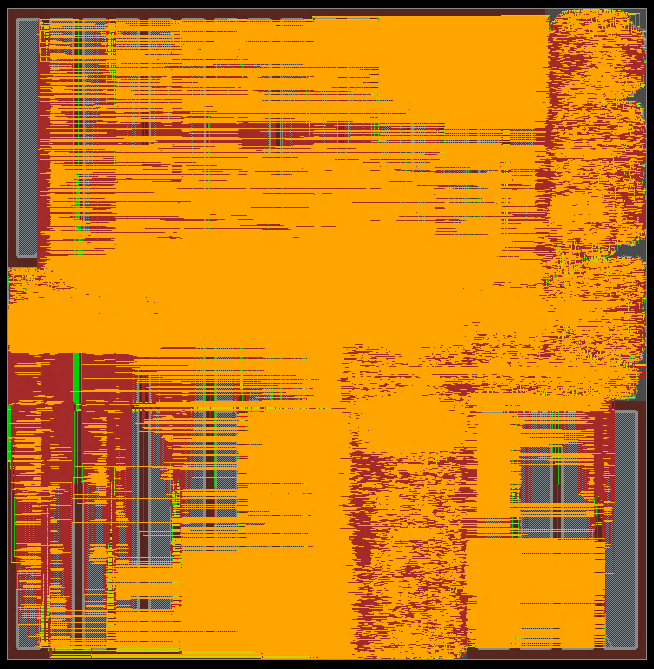}
  \caption{GDS Layout}
  \label{fig:DieWires}
\end{subfigure}
\begin{subfigure}{.48\columnwidth}
\centering
  \includegraphics[width=0.65\columnwidth]{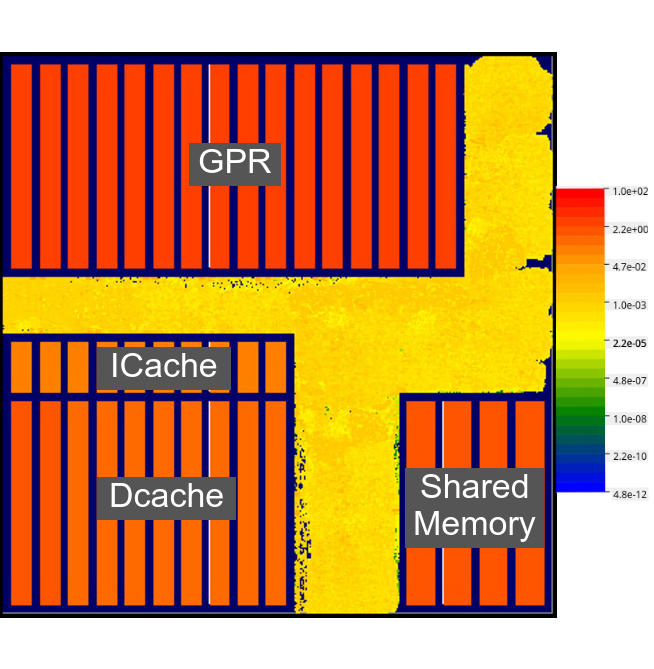}
  \caption{Power density distribution across Vortex}
  \label{fig:DiePower}
\end{subfigure}
\caption{GDS layouts for our Vortex with 8 warp, 4 thread configuration (4KB register file). The design was synthesized for 300Mhz and produced a total power output of 46.8mW. 4KB 2 ways 4 banks-data cache, 8KB with 4 banks-shared memory, 1Kb 2 way cache, one bank-I cache.}
\label{fig:die}
\end{figure}

         

\section{Evaluation}

This section will evaluate both the RTL Verilog model for Vortex and the software stack. 

\subsection{Micro-architecture Design space explorations}

In Vortex design, we can increase the data-level parallelism either by increasing the number of threads or by increasing the number of warps. Increasing the number of threads is similar to increasing the SIMD width and involves the following changes to the hardware: 1) Increasing the GPR memory width for reads and writes, 2) Increasing the number of ALUs to match the number of threads, 3) increasing the register width for every pipeline stage after the GPR read stage, 4) increasing the arbitration logic required in both the cache and the shared memory to detect bank conflicts and handle cache misses, and 5) increasing the number of IPDOM enteries. 

Whereas, increasing the number of warps does not require increasing the number of ALUs because the ALUs are multiplexed by a higher number of warps. Increasing the number of warps involves the following changes to the hardware: 1) increasing the logic for the warp scheduler, 2) increasing the number of GPR tables, 3) increasing the number of IPDOM stacks, 4) increasing the number of register scoreboards, and 5) increasing the size of the warp table. It's important to note that the cost of increasing the number of warps is dependant on the number of threads in that warp; thus increasing warps for bigger thread configurations becomes more expensive. This is because the size of each GPR table, IPDOM stack, and warp table are dependant on the number of threads.

\begin{figure}
    \centering
    \includegraphics[width=0.8\columnwidth]{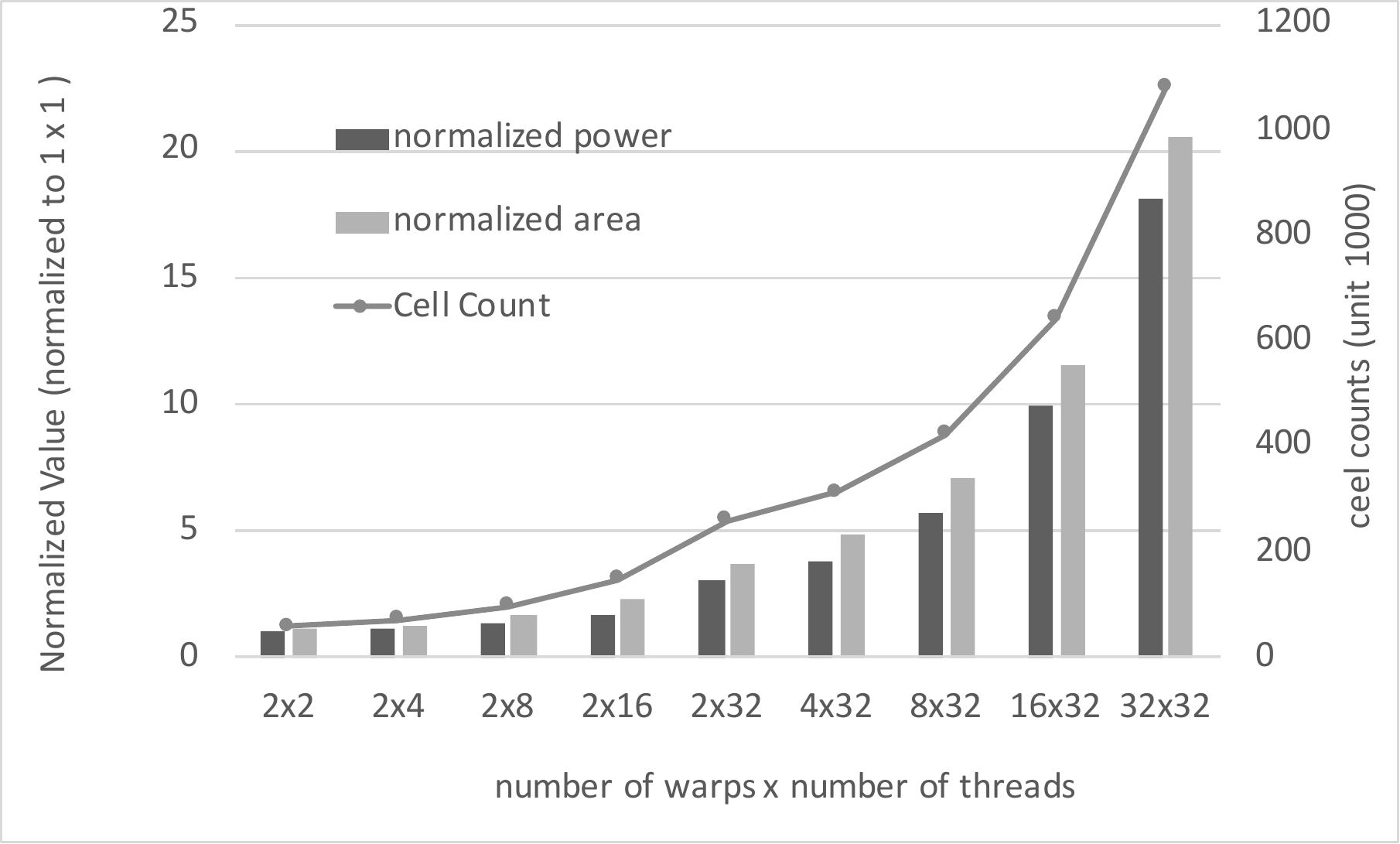}
    \caption{Synthesized results for power, area and cell counts for different number of warps and threads }
    \label{fig:area}
\end{figure}

Figure~\ref{fig:area} shows the increases in the area and power as we increase the number of threads and warps. The number is normalized to 1 warp and 1 thread support. All the data includes 1Kb 2 way instruction cache, 4 Kb 2 way 4 banks data cache, and an 8kb 4 banks shared memory module.

\subsection{Benchmarks}

All the benchmarks used for evaluations were taken from the Rodinia  \cite{rodinia_bench}, a popular GPGPU benchmark suite.\footnote{The benchmarks that are not evaluated in this paper is due to the lack of support from LLVM RISC-V.} 


\subsection{simX Simulator}

Because the benchmarks used in Rodinia Benchmark Suite have large data-sets that took Modelsim a long time to simulate, we used simX, a C++ cycle-level in-house simulator for Vortex with a cycle accuracy within 6\% of the actual Verilog model. Please note that power and area numbers are synthesized from the RTL. 

\subsection{Performance Evaluations}

\begin{figure}
    \centering
    \includegraphics[width=\columnwidth]{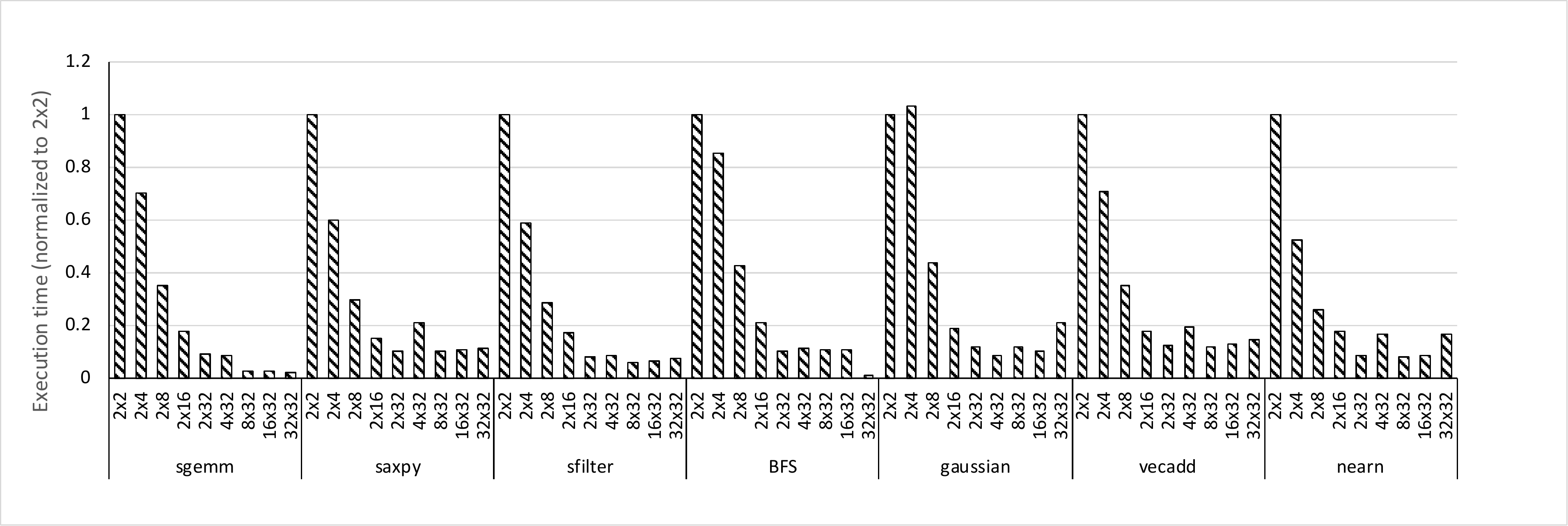}
    \caption{Performance of a subset of Rodinia benchmark suite (\# of warps x \# of threads}
    \label{fig:perf_mini}
\end{figure}

Figure~\ref{fig:perf_mini} shows the normalized execution time of the benchmarks normalized to the 2warps x  2threads configuration. As we predict, most of the time, as we increase the number of threads (i.e., increasing the SIMD width.), the performance is improved, but not too much from increasing the number of warps.  Some benchmarks get benefits from increasing the number of warps such as bfs, but in most of the cases increasing the number of warps is not translated into performance benefit.
The main reason is that to reduce the simulation time, we warmed up caches and reduced the data set size, thereby the cache hit rate in the evaluated benchmarks was high. Increasing the number of warps is typically useful to hide long latency operations such as cache misses by increasing TLP and MLP; Thus, the benchmark that benefited the most from the high warp count is BFS which is an irregular benchmark.

\begin{figure}
    \centering
    \includegraphics[width=\columnwidth]{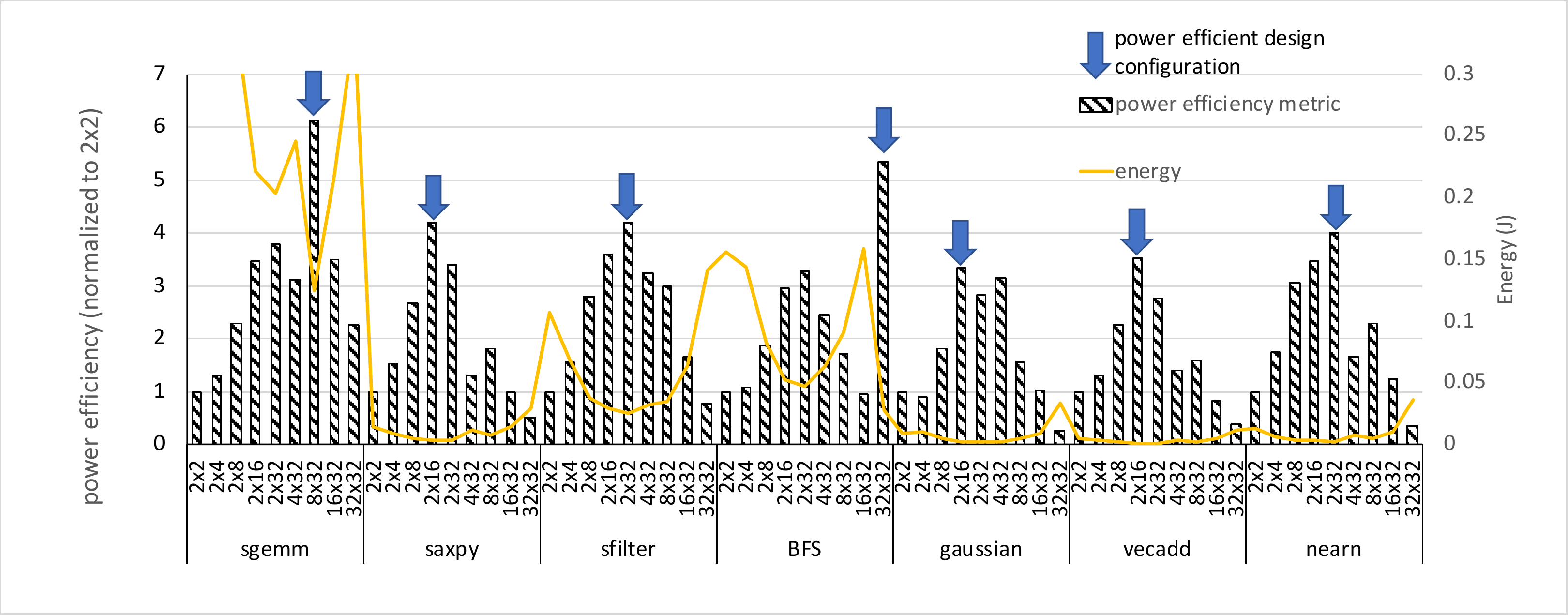}
    \caption{Power efficiency (\# of warps x \# of threads (Power efficiency and Energy)}
    \label{fig:perf_power_mini}
\end{figure}

As we increase the number of threads and warps, the power consumption increases but they do not necessarily produce more performance. Hence, the most power efficient design points vary depending on the benchmark. 
Figure~\ref{fig:perf_power_mini} shows a power efficiency metric (similar to performance per watt) which is normalized to the 2 warps x 2 threads configuration. 
The results show that for many benchmarks, the most power efficient design is the one with fewer number of warps and 32 threads except for the BFS benchmark. As we discussed earlier since BFS benchmark gets the best performance from the 32 warps x 32 threads configuration, it also shows the most power efficient design point. 

\subsection{Placement and layout} 

We synthesized our RTL using a 15-nm educational library. Using Innovus, we also performed Place and route (PnR). Figure~\ref{fig:die} shows the GDS layout and the power density map of our Vortex processor. From the power density map, we observe that the power is well distributed among the cell area. In addition, we observe the memory including the GPR, data cache, instruction icache and the shared memory have a higher power consumption. 


\section{Related Work}
\label{sec:related}

ARA \cite{ARA} is a RISC-V Vector Processor that implements a variable-length single-instruction multiple-data execution model where vector instructions are streamed into vector lanes and their execution is time-multiplexed over the shared execution units to increase energy efficiency. Ara design is based on the open-source RISC-V Vector ISA Extension proposal \cite{RISCVV} taking advantage of its vector-length agnostic ISA and its relaxed architectural vector registers. Maximizing the utilization of the vector processors can be challenging, specifically when dealing with data dependent control flow. That is where SIMT architectures like Vortex present an advantage with their flexible scalar-threads that can diverge independently. 


HWACHA \cite{HWACHA} is a RISC-V scalar processor with a vector accelerator that implements a SIMT-like architecture where vector arithmetic instructions are expanded into micro-ops and scheduled on separate processing threads on the accelerator. An advantage that Hwacha has over pure SIMT processors like Vortex is its ability to overlap the execution of scalar instructions on the scalar processor which increases hardware complexity for hazard management.

Simty \cite{collange2017simty} processor implements a specialized RISC-V architecture that supports SIMT execution similar to Vortex, but with different control flow divergence handling. In their work, only microarchitecture was implemented as a proof of concept and there was no software stack, and none of GPGPU applications were executed with the architecture.  


\section{Conclusions}
\label{sec:conclusions}

In this paper we proposed Vortex that supports an extended version of RISC-V  for GPGPU applications. We also modified OpenCL software stack (POCL) to run various OpenCL kernels and demonstrated that. We plan to release the Vortex RTL and POCL modifications to the public.\footnote{currently the github is private for blind review.} We believe that an Open Source version of RISC-V GPGPU will enrich the RISC-V echo systems and accelerate other researchers that study GPGPUs in wider topics since the entire software stack is also based on Open Source implementations. 


\bibliographystyle{IEEEtranS}
\bibliography{ref,proposal}

\end{document}